\documentclass[prb,twocolumn,showpacs]{revtex4}
\usepackage{graphicx}
\usepackage{amsmath}
\usepackage{amssymb}

\renewcommand{\Im}{\mathrm{Im}\,}

\renewcommand{\Im}{\mathrm{Im}\,}
\newcommand{\bsigma}{\boldsymbol{\sigma}}
\DeclareMathAlphabet{\bi}{OML}{cmm}{b}{it}

\begin{document}
\title{Thermodynamic properties of a magnetically modulated graphene}
\author{SK Firoz Islam, Naveen K. Singh and Tarun Kanti Ghosh}
\affiliation{Department of Physics. Indian Institute of Technology-Kanpur, 
Kanpur-208 016, India}

\begin{abstract}
The effect of magnetic modulation on thermodynamic properties of
a graphene monolayer in presence of a constant perpendicular magnetic 
field is reported here.  One-dimensional spatial electric or magnetic 
modulation lifts the degeneracy of the Landau levels and converts  
into bands and their bandwidth oscillates with magnetic field leading 
to Weiss-type oscillation in the thermodynamic properties. 
The effect of magnetic modulation on thermodynamic properties of a 
graphene sheet is studied and then compared with electrically modulated 
graphene and magnetically modulated conventional two-dimensional electron 
gas (2DEG). We observe Weiss-type and de Haas-van Alphen (dHvA) oscillations 
at low and high magnetic field, respectively.
There is a definite phase difference in Weiss-type 
oscillations in thermodynamic quantities of magnetically modulated 
graphene compared to electrically modulated graphene.
On the other hand, the phase remains same and amplitude of the oscillation 
is large when compared with the magnetically modulated 2DEG.
Explicit asymptotic expressions of density of states and the 
Helmholtz free energy are provided to understand the phase and amplitude 
of the Weiss-type oscillations qualitatively. 
We also study thermodynamic properties when both electric and
magnetic modulations are present. The Weiss-type oscillations still exist when
the modulations are out-of-phase. 
\end{abstract}

\pacs{65.80.+n,71.70.Di,71.18.+y}



\date{\today}
\maketitle

\section{Introduction}
Graphene is a two-dimensional sheet of carbon allotrope with
honeycomb lattice structure. It can be considered as a basic 
building block of all other dimensional carbon allotrope 
\cite{neto,klein,zhang,li}. 
The relativistic-like, massless and linear energy spectrum of graphene's 
quasi-particles in low-energy range close to the Dirac points in it's 
band structure reflect in it's different properties like transport 
properties, optical properties, magnetic properties etc in different way 
than the conventional 2DEG formed in the semiconductor heterostructures. 
The massless linear energy dispersion and charge conjugation symmetry 
cause some unusual phenomena like Klein paradox, anomalous quantum 
Hall effect and non-zero Berry phase \cite{novo,geim,kats,ando,shara}.

Effect of electric or/and magnetic modulations on transport and 
thermodynamic properties of quantum 2DEG systems is continuing 
to be an active research field. 
The magnetotransport properties of a conventional 2DEG in presence of 
a one-dimensional (1D) weak electric modulation have been studied
in great details experimentally and theoretically by various groups 
\cite{weiss,ploog,peter,ger,been,beat}. Later, the magnetotransport
properties of a magnetically modulated 2DEG in presence of a constant
perpendicular magnetic field have been studied theoretically 
\cite{fm,vasi,physica,peng} and also experimentally \cite{exp1,exp2,exp3}. 
The presence of weak electric/magnetic modulation broadens 
the Landau energy levels into bands. The band width oscillates with 
the magnetic field and its oscillations are also reflected in 
magnetotransport properties. It has been observed that the
magnetoresistivity tensor oscillates with inverse of the magnetic 
field at very low magnetic field. This oscillation is known as the 
Weiss oscillation which is completely different from the 
Shubnikov-de Hass (SdH) oscillations observed at higher magnetic field. 
Period of Weiss oscillation varies with electron 
density $n_e$ as $\sqrt n_e$, whereas for SdH it depends linearly on $n_e$. 
The Weiss oscillation is due to effect of the commensurability between two
length scales in the system: the cyclotron diameter at the
Fermi energy and the period of the modulated electric/magnetic potential.
Magnetotransport properties of electrically \cite{matuli,tahir}
and magnetically modulated graphene \cite{sabeeh} in presence of a constant
perpendicular magnetic field have been studied recently.

The effect of weak electric and magnetic modulation on thermodynamic 
properties for 2DEG in presence of a perpendicular magnetic field have 
been studied theoretically \cite{peter,stewart,tong}. It is observed 
that the Weiss-type oscillation in various thermodynamic properties 
are present. The Weiss-type oscillations are completely different in origin 
from the usual de Haas-van Alphen (dHvA) oscillations which occurs
at high magnetic field. The dHvA oscillation is effect of the formation of 
discrete Landau energy levels due to the quantizing magnetic field.
Recently, thermodynamic properties of a monolayer graphene in presence of
weak electric modulation have been studied and the Weiss-type oscillations 
are seen \cite{khan}. These results inspired us to study thermodynamic 
properties of a magnetically modulated graphene sheet in presence of a 
constant magnetic field.

The source of the magnetic modulation, magnetic stripes or superconducting
materials, acts like electrical gates and produces an electric modulation.
The transport properties of a 2DEG \cite{fm} and a graphene sheet 
\cite{ijmp} in presence of both the modulations were studied.

In this paper we study the effect of magnetic modulation on thermodynamic 
properties of a graphene sheet in presence of a constant magnetic field. 
We compare our results with electrically modulated graphene and with 
magnetically modulated conventional 2DEG. We also calculate an asymptotic 
expression of density of states (DOS) and the Helmholtz free energy of 
a magnetically modulated graphene in presence of a constant magnetic 
field at low temperature. In addition to these, we also study thermodynamic
properties of graphene and 2DEG when both electric and magnetic modulations 
are present.

This paper is organized as follows. In section II, we summarize 
the standard results of the energy spectrums and the corresponding 
eigenstates for electrically and magnetically modulated graphene 
layer and 2DEG in presence of a constant magnetic field.  
In section III, we numerically calculate the thermodynamic 
quantities like, Helmholtz free energy, internal energy, entropy, 
heat capacity and magnetization. We analyze our numerical results 
and compare with electrically modulated graphene and magnetically 
modulated 2DEG in section IV. We discuss the behavior of the 
asymptotic expression of the DOS and the Helmholtz free energy
for magnetically modulated graphene in section V.
In section VI, we study thermodynamic properties of graphene and 2DEG 
in presence of both electric and magnetic modulations.
We present summary of our work in section VII. The detail calculation 
of the DOS by using Green's function method is presented in the Appendix 1. 
 
\section{Energy spectrum}
We consider a monolayer graphene sheet subjected to a 
perpendicular constant magnetic field ${\bf B}_0 = B_{0}\hat{z}$, the 
Hamiltonian of an electron with charge $-e$ is given by
\begin{eqnarray}
H_0^G & = & v_{_F} \bsigma \cdot ({\bf p} + e {\bf A}_0)
\end{eqnarray}
where $ \bsigma = (\sigma_x,\sigma_y)$ are the Pauli matrices, 
$v_{_F} \approx 10^6 $ m/s 
is the Fermi velocity and ${\bf A}_0$ is the magnetic vector potential. 
Here, we have chosen the Landau gauge ${\bf A}_0 = B_{0}x \hat y$.
The energy eigenvalues are $ E_{n}^{g} = \hbar \omega_g \sqrt{2n} $,
where $n=0,1,2,3...$ is the Landau level index and  
$\omega_g = v_{_F} \sqrt{eB_{0}/\hbar}$ is the cyclotron frequency. 
The corresponding normalized eigenstates are
\begin{equation}
\Psi_{n,k_y}(x,y) = \frac{e^{ik_y y}}{\sqrt{2L_y l_0}}
\begin{bmatrix}
-i\phi_{n-1}[(x+x_0)/l_0]\\\phi_{n}[(x+x_0)/l_0]
\end{bmatrix},
\end{equation}
where $ \phi_{n}(x) $ is the known harmonic oscillator
wave function,  
$l_0 = \sqrt{\hbar/(eB_{0})}$ is the magnetic length, 
$x_0 = - k_y l_0^2$ is the center of the cyclotron orbit 
and $L_y$ is the width of the graphene in the $y$-direction.

We consider the perpendicular magnetic field is modulated very 
weakly by ${\bf B_{1} } = B_{1} \cos{(qx)} \hat{z}$,
where $q=2\pi/a$ and $a$ is the modulation period. 
The total Hamiltonian can be written as $ H = H_0^{G} + H_1 $, where
$ H_1 = V_m^g \sigma_y \sin{(qx)} $. Here,
with $ V_m^g = e v_{_F} B_1/q $ is the strength of the effective
magnetic potential determined by the amplitude $B_1$ 
and the period $a$ of the magnetic modulation.
Using the first-order perturbation theory, the energy 
correction to the unperturbed Hamiltonian $H_0^G$ is given 
as \cite{sabeeh}
\begin{eqnarray}
\Delta E_{n,k_y}^{g,m} & = & V_m^g \sqrt{\frac{n}{u}}e^{-u/2}
[L_{n-1}(u)-L_n(u)]\cos(qx_0),
\end{eqnarray}
where $L_n(u)$ is the Laguerre 
polynomial and $ u = q^2l_0^2/2$.
So the total energy upto the first-order in $V_m^g$ is given by
$ E_{n,k_y}^{g,m} = E_n^g + \Delta E_{n,k_y}^{g,m} $.
The band width in presence of the magnetic modulation is
$ \Delta_m \sim \sin(2 \sqrt{nu} - \pi/4) $. Using the flat band
condition, we get $ 2R_c = a (j + 1/4) $, with $j = 0,1,2,3,...$ and
$R_c = k_{_F} l_0^2$.

The energy correction due to the weak electric modulation 
$U(x)=V_e^g \cos{(qx)}$ can be obtained by the same method 
and it is given as \cite{matuli}
\begin{equation}
\Delta E_{n,k_y}^{g,e} = \frac{V_e^g}{2}e^{-u/2} 
\left[L_{n}(u)+L_{n-1}(u) \right]\cos(qx_0).
\end{equation}
So the total energy upto the order of $V_e^g$ is
$ E_{n,k_y}^{g,e} = E_n^g + \Delta E_{n,k_y}^{g,e} $.
The bandwidth in presence of the electric modulation is
$\Delta_e \sim \cos(2 \sqrt{n u} - \pi/4)$.
The bandwidths due to magnetic modulation and due to the 
electric modulation are out of phase.
The band will be flat when 
$2R_c = a (j + 3/4) $ which is different from what we get
in the magnetic modulation case.

The Hamiltonian of a conventional 2DEG in presence of a perpendicular 
constant magnetic field ${\bf B_{0}} $ is
\begin{equation}
H_0^{2d} = \frac{p_x^2}{2m^*} + \frac{1}{2m^*} (p_y + e A_y)^2.
\end{equation}
The energy spectrum is $E_{n}^{2d} = \hbar\omega_c (n + 1/2)$,
where $ n = 0, 1, 2, 3....$ and $\omega_0 = eB_{0}/m^* $ is the 
cyclotron frequency. The corresponding eigenstates are
\begin{equation}
\Psi_{n,k_y}(x,y) = \frac{e^{ik_y y}}{\sqrt{L_y l_0}}
\phi_{n}[(x+x_0)/l_0].
\end{equation}
In presence of the weak magnetic modulation $ {\bf B}_1$, the 
total Hamiltonian is $ H = H_0^{2d} + H_1^{2d} $, where
$ H_1^{2d} =[V_m^{2d}/(\hbar q)](p_{y} + e B_0 x)\sin(qx) $
with $V_m^{2d}=\hbar (eB_{1}/m^*)$ is the strength of the effective 
magnetic potential determined by the amplitude of the magnetic modulation.
The first-order energy correction due to the weak magnetic modulation 
is given by \cite{vasi,fm}
\begin{eqnarray}
\Delta E_{n,k_y}^{2d,m}  &=&  V_m^{2d} e^{-\frac{u}{2}} 
\Big[ \frac{u -2n}{2u} L_n(u)\nonumber\\
&+&\frac{n}{u}L_{n-1}(u) \Big] \cos(qx_0).
\end{eqnarray}
The total energy of a 2DEG in presence of the magnetic modulation is
then $ E_{n,k_y}^{2d,m} = E_{n}^{2d} + \Delta E_{n,k_y}^{2d,m} $.
The energy correction due to electric modulation is given by
\begin{equation}
 \Delta E_{n,k_y}^{2d,e} = \frac{V_e^{2d}}{2}e^{-u/2} 
\left[L_{n}(u)+L_{n+1}(u) \right]\cos(qx_0).
\end{equation}

All these standard results will be used to calculate the thermodynamic 
properties numerically in the next section.

\section{Thermodynamic quantities}
In this section we discuss all standard thermodynamic equations 
to be used for calculating chemical potential, Helmholtz free energy, 
internal energy, entropy, magnetization and specific heat.

The DOS of a magnetically modulated graphene sheet in presence of a
constant magnetic field can be written as
\begin{equation}
D(E)=\frac{A}{\pi l_0^2} \sum_{n,k_y} \delta(E - E_{n,k_y}),
\end{equation}
where $A = L_x L_y$ is the area of the graphene sheet
and $E_{n,k_y} $ is a energy dispersion of a given system
like $ E_{n,k_y}^{g,m}, E_{n,k_y}^{g,e}, E_{n,k_y}^{2d,m}$. 
The dependence of chemical potential $\mu(B,T)$ on 
temperature ($T$) and magnetic field $(B)$ can be obtained 
numerically by using the following normalization condition
\begin{equation}
N = \int_0^{\infty} D(E) f(E) dE,
\end{equation}
where $N$ is the total number of electrons, 
$f(E) = [exp(\frac{E-\mu}{k_B T}) + 1]^{-1}$ is the 
Fermi-Dirac distribution function and $k_{_B}$ is the 
Boltzmann constant. Using the expression of $D(E)$ given 
in the above equation we get
\begin{equation}
n_e \pi l_0^2 = \frac{1}{\pi} \sum_{n=0}^{\infty} 
\int_0^\pi dt f(E_{n,t}),
\end{equation}
where $n_e$ is the electron density and $t=qx_0$.

The total internal energy can be written as
\begin{equation}
U = \int_0^{\infty} E D(E)f(E) dE.
\end{equation}
The internal energy per unit area is
\begin{equation}
\frac{U}{A}=\frac{1}{\pi^2 l_0^2} \sum_{n=0}^{\infty} 
\int_0^\pi E_{n,t} f(E_{n,t})dt.
\end{equation}
Now for a system of non-interacting electrons, the Helmholtz 
free energy density \cite{pat} is given by
\begin{equation}
\frac{F}{A} = \mu n_e - \frac{k_{_B} T}{\pi}
\sum_{n=0}^{\infty} \int_0^\pi 
\ln\left[1 + exp\left(\frac{\mu - E_{n,t}}{k_{_B} T}\right)\right]dt.
\end{equation}
From the above equations it is clear that the DOS plays 
an important role in the behavior of the thermodynamic 
properties. In presence of the perpendicular magnetic field 
the DOS shows a series of delta function because of the 
quantized energy spectrum. Graphene and conventional 2DEG 
having different energy spectrums reflects differently in their 
thermodynamic properties.
By using the above results we compute entropy via $S = (U-F)/T$, 
orbital magnetization $M = - (\partial F/\partial B_0)_{A,N}$ and 
heat capacity $C = T (\partial S/\partial T)_{A,N}
= - T (\partial^2 F/\partial^2 B_0)_{A,N} $. 
For better visualization of effect of the magnetic modulation 
we plot the fluctuation $\Delta \Pi = \Pi(B_1)-\Pi(B_1 = 0)$, 
where $B_1$ is the strength of magnetic modulation, and 
$\Pi$ is a thermodynamic quantity like $\mu, F, U, M, S, C $.

\section{Numerical Results and Discussions}
Thermodynamic properties of magnetically modulated graphene sheet 
in presence of a constant magnetic field are studied. The aim is 
to study the effect of magnetic modulation on graphene in comparison 
with the electrically modulated graphene and magnetically modulated 
conventional 2DEG. We plot the fluctuation due to weak modulation 
in chemical potential, Helmholtz free energy, magnetization, internal 
energy, entropy and specific heat with the magnetic field $B_{0}$.
We have used the following parameters for numerical calculation: 
electron density $n_e$ = 3.16 $\times 10^{15}/$ m${}^2$, 
effective mass of an electron $m^* = 0.067 m_e$
with $m_e$ is the bare electron mass, temperature $T=2$ K, 
modulation period $a = 382 $ nm as used in \cite{khan,stewart}. 
For these parameters, $ E_F^{g} =0.1 $ eV and $ E_F^{2d} =14.25 $ meV.

\begin{figure}[ht]
\begin{center}\leavevmode
\includegraphics[width=95mm]{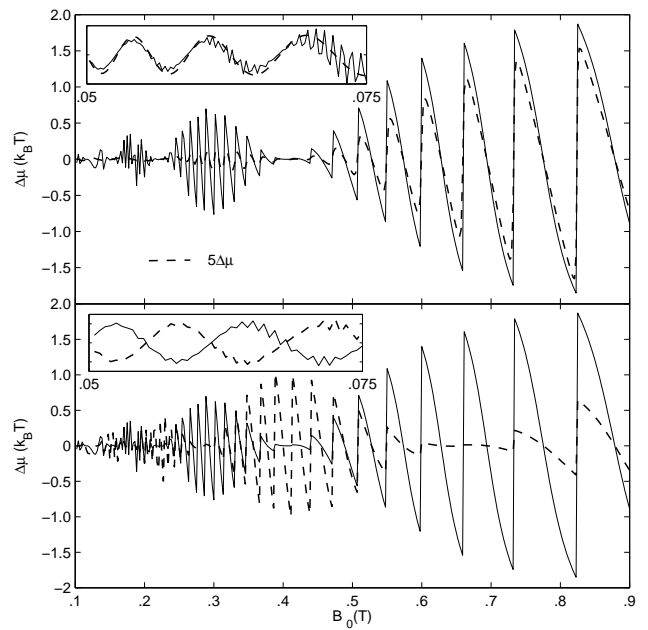}
\caption{Plots of the fluctuation in chemical potential
vs magnetic field at $T=2 $ K.}
\label{Fig1}
\end{center}
\end{figure}

\begin{figure}[ht]
\begin{center}\leavevmode
\includegraphics[width=95mm]{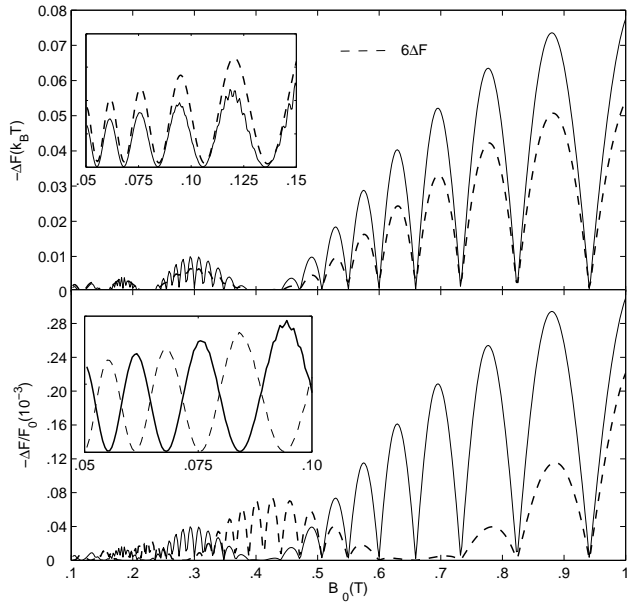}
\caption{Plots of the change in the free energy vs magnetic field.
To make the fluctuation dimensionless we use $ F_0 = E_F^{2d}/2 $ in
the lower panel but as $E_{F}$ is different for the two systems
we are showing fluctuation in units of $k_{_B} T$ in the upper panel.}
\label{Fig2}
\end{center}
\end{figure}

In figures 1-6, we have plotted the fluctuations in various 
thermodynamic properties, $ \Delta \Pi $, due to both magnetic and 
electric modulations. We have scaled all the thermodynamic quantities 
per electron. 
Each figure contains two panels, the upper panel shows the effect 
of the magnetic modulation on graphene (solid line) and 
conventional 2DEG (dashed line), and 
the lower panel shows this fluctuation for the electrically modulated 
(dashed line) 
and magnetically modulated (solid line) graphene. 
We have zoomed the oscillations at low magnetic field and shown in 
the inset of figures 1-6.
It clearly shows 
that weak 1D periodic potential, either electric or magnetic in nature,
induces new oscillations at low magnetic field. These modulation induced 
oscillation is due to the commensurability of the two length scales 
present in the system. These oscillations are similar to the Weiss 
oscillations observed in the magnetoresistance at low magnetic fields.

It is clear from the upper panels of the figures 1-6 that the 
fluctuation in the thermodynamic properties, $\Delta \Pi$, of 
the graphene sheet is quite large compared to the 2DEG system.
This can be understood qualitatively from the following arguments.
The energy correction for the magnetically modulated
graphene and 2DEG systems are $V_m^g = 1$ meV
and $V_m^{2d} = 0.046$ meV for the same
strength of the magnetic modulation $B_{1}=0.02$ T.
Clearly, the energy correction due to the magnetic modulation in graphene 
is quite large compared to that of the 2DEG. This is the origin for higher 
amplitude fluctuation of the Weiss-type and dHvA oscillations in graphene 
compared to the 2DEG.
It is interesting to note that amplitude of diffusive
conductivity in magnetically modulated graphene is small
compared to magnetically modulated 2DEG \cite{sabeeh}.

The lower panels of the figures 1-6 show that the Weiss-type oscillation 
has a definite phase difference in the thermodynamic properties 
between magnetic and electric modulation cases, which is due to the
following reasons. The phase difference in the fluctuations between 
electrically and magnetically modulated graphene comes from the nature 
of the energy correction. In the electric modulation case the energy 
correction contains addition of the two successive Laguerre polynomials 
and in magnetic modulation case it is subtraction of the two 
successive Laguerre polynomials, giving rise to the cosine and sine 
term, respectively.
We also observe that the dHvA-type oscillations for 
the two different kind of modulations remain in the same phase 
with each other.
The phase difference between the two different kind of modulated systems
are shown explicitly by using the analytical expressions of the DOS in the
next section.

Figures (1) and (3) shows the fluctuation in chemical potential 
and magnetization with magnetic field, respectively.
The fluctuation in magnetically modulated graphene is
several times higher than the magnetically modulated 
2DEG in the Weiss-type oscillation, but their phases remain
same.
On the other hand, the magnetically modulated graphene 
shows $\pi$ phase difference in Weiss-type oscillation 
compared to the electrically modulated graphene with 
the same amplitude.

Figure (2) is showing the fluctuation in the Helmholtz free 
energy where lower panel shows $\pi/2$ phase differences between 
electric and magnetic modulation cases in graphene with the 
same amplitude.
The amplitude of fluctuation of graphene is several times higher 
than the 2DEG but the phases remain same.
From the flat-band condition, the minima of the
bandwidth occur at $ B(T) = 0.092, 0.113, 0.148, 0.214, 0.386 $. 
On the other hand, the fluctuation in the free energy vanishes at 
$B(T) = 0.092, 0.115, 0.150, 0.214, 0.385 $. It shows that the
minima of the free energy fluctuation occur at those values
of the magnetic field where the bandwidth minima occur.

We have plotted internal energy fluctuation in Fig. (4). 
The lower panel of Fig. (4) shows that the Weiss-type 
oscillation is appeared with same amplitude but 
$\pi/2$ phase difference when compared with the electrical 
modulation case. When we compare with conventional 2DEG,
it is similar to the case of the Helmholtz free energy i. e.
amplitude is higher in magnitude.

\begin{figure}[ht]
\begin{center}\leavevmode
\includegraphics[width=95mm]{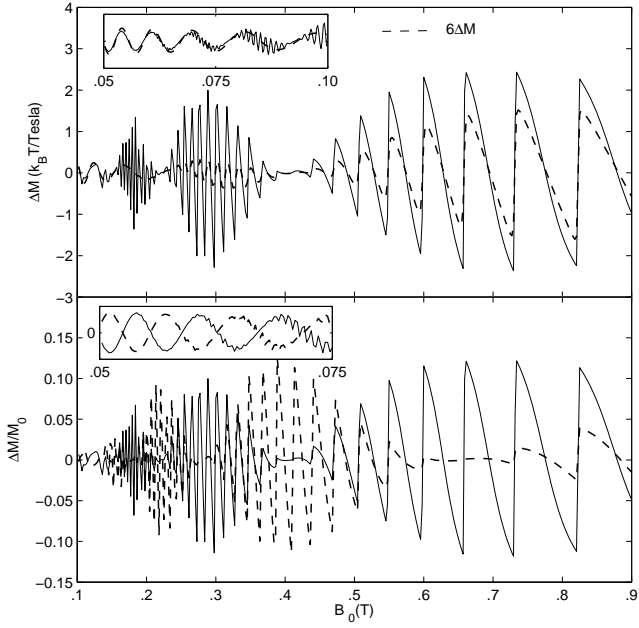}
\caption{Plots of the fluctuation in magnetization vs magnetic field.
In the lower panel $\Delta M$ is scaled by $ M_0 =  \mu^{\ast}_{_B}$, 
where $\mu_{_B}^{\ast} = e \hbar/ 2m_g $ is the effective Bohr magneton
with $ m_g = E_F^g/v_{_F}^2$ is the cyclotron mass.}
\label{Fig3}
\end{center}
\end{figure}

Figures (5) and (6) are showing the entropy and specific heat 
fluctuation, respectively. The fluctuation in graphene is
higher by several times than that of the 2DEG.
The phase relationship is not discernible in these figures.
 
In all the above cases, dHvA-type oscillation remains in the same 
phase and does not depend on the modulation type as it
is the manifestation of the quantized Landau levels rather 
than periodic perturbation.
Figure (7) shows the damping of fluctuation in chemical potential 
and free energy with increasing temperature. Temperature dependence 
in Weiss-type and dHvA oscillations are independent of type of modulation 
and already discussed in electrical modulation case. 
The phase relationships for all the thermodynamic quantities are given in 
Table (1).
\begin{table}[ht]
\centering
\begin{tabular}{|c |c |c | c| c| c| c|}
\hline\hline
Case & $ \Delta \mu $ & $ \Delta F$  & $ \Delta M $ & 
$ \Delta U$ & $ \Delta S$ & $ \Delta C$  \\
\hline
Phase shift & $\pi$ & $\pi/2$ & $\pi$ & $\pi/2 $ &  
indiscernible &  indiscernible\\
\hline
\end{tabular}
\caption{The phase shifts in the Weiss-type oscillations appear in
the fluctuation of thermodynamic quantities between the electrically
and magnetically modulated  graphene.}
\end{table}

Even at higher magnetic field, the effect of modulation
on the fluctuation of thermodynamic quantities is still 
exist which is the manifestation of the modulated density 
of states. The dHvA-type oscillation is corresponding to 
the crossing of each Landau level one by one through 
the Fermi level. The density of available states 
per Landau level is $1/2\pi l_0^2 $ and 
it increases linearly with magnetic field which results
in increasing amplitude of dHvA oscillation with magnetic field.

\begin{figure}[ht]
\begin{center}\leavevmode
\includegraphics[width=95mm]{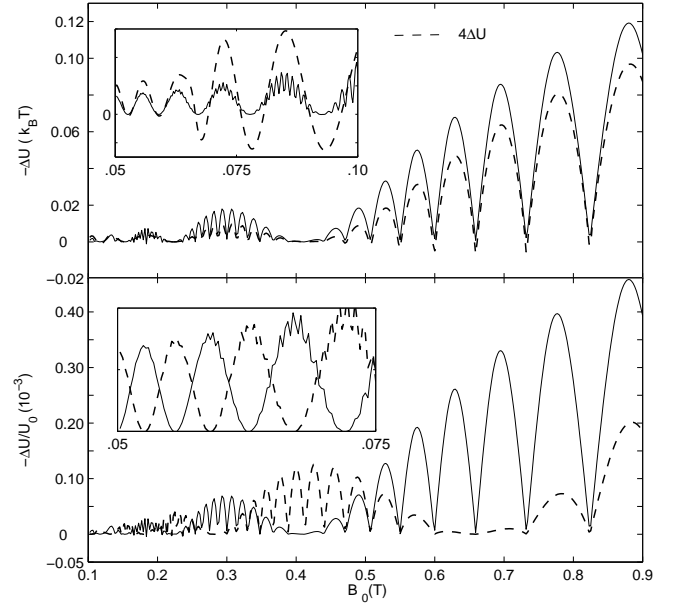}
\caption{Plots of the fluctuation in the internal energy vs magnetic field.
In the lower panel $\Delta U$ is scaled by $U_0=E_F^{g}/2$.}
\label{Fig4}
\end{center}
\end{figure}

\begin{figure}[ht]
\begin{center}\leavevmode
\includegraphics[width=95mm]{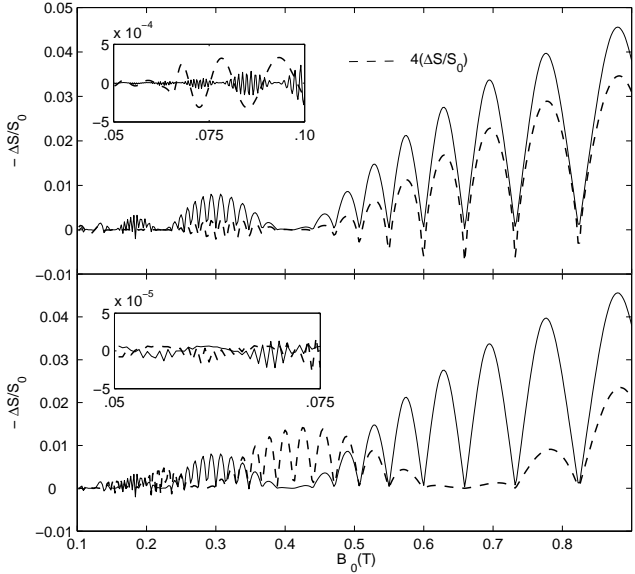}
\caption{Plots of the fluctuation in entropy vs magnetic field.  
In the lower panel $ \Delta S $ is scaled by $S_0 =k_{_B}$. }
\label{Fig5}
\end{center}
\end{figure}

\begin{figure}[ht]
\begin{center}\leavevmode
\includegraphics[width=95mm]{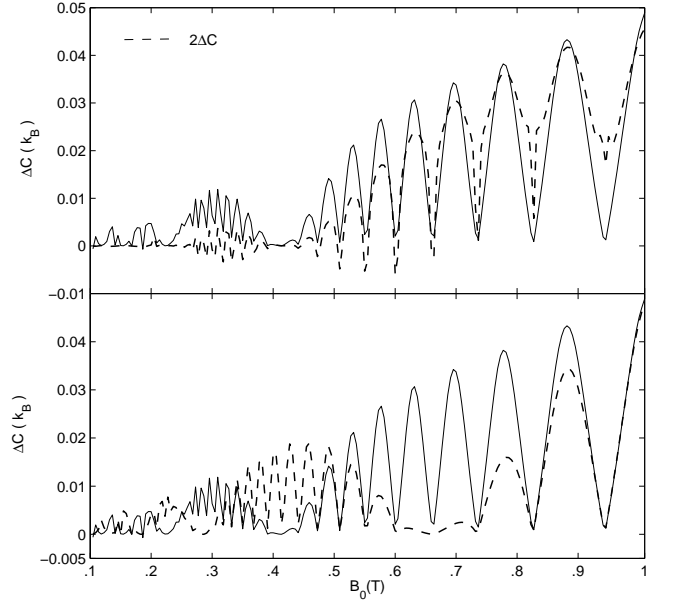}
\caption{Plots of the fluctuation in specific heat vs magnetic field.}
\label{Fig6}
\end{center}
\end{figure}

\begin{figure}[ht]
\begin{center}\leavevmode
\includegraphics[width=95mm]{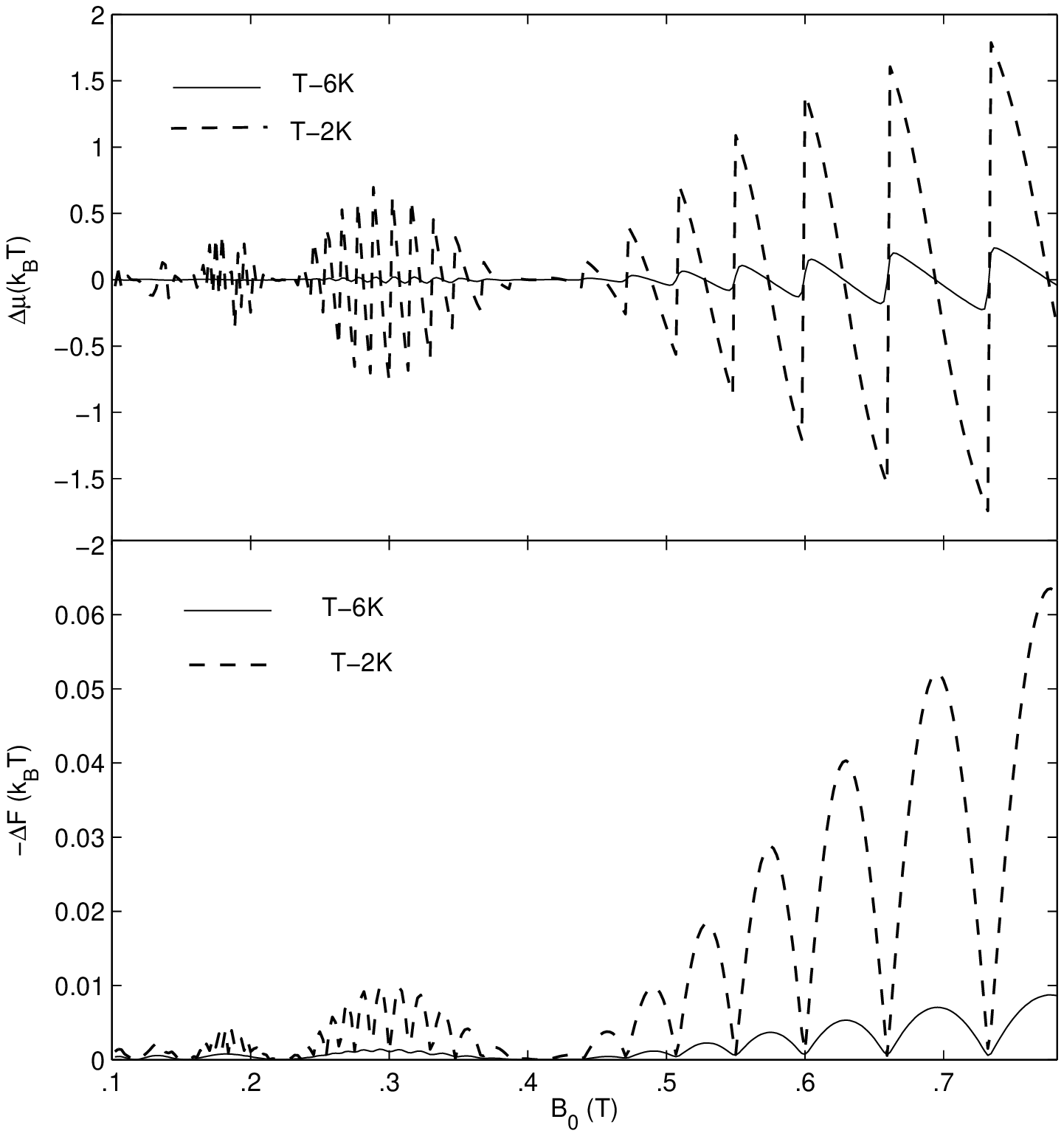}
\caption{Plots of the fluctuation in chemical potential and free energy vs 
magnetic field for $T=2$ K and $T= 6$ K with the same parameter.}
\label{Fig7}
\end{center}
\end{figure}

\section{Asymptotic Results}
Here we derive an asymptotic expression for the DOS and the Helmholtz 
free energy.
For weak magnetic modulation and under quasi-classical limit,
we calculate the DOS by using the Green's function technique 
(see the Appendix 1) and written as a sum of the modulated and
unmodulated part as 
$ D(\epsilon) = D_u(\epsilon) + D_m(\epsilon) \label{dos} $,
where
\begin{equation}
D_u(\epsilon)=\frac{1}{\pi l_0^2}\frac{\epsilon}{\hbar\omega_g}
\Big[1 + 2 \cos\left(\pi \epsilon^2 \right)\Big],\label{du}
\end{equation}
\begin{equation}
D_m(\epsilon)  =  - \frac{2\Omega_m}{\pi l_0^2} \frac{\epsilon}{\hbar\omega_g}
\Big[ \epsilon^3 \cos \left( \pi \epsilon^2 \right)
 \sin^2\left(ql_0 \epsilon
-\frac{\pi}{4}\right)\Big]\label{dm},
\end{equation}
and
\begin{equation}
\Omega_m = \frac{(V_m^g)^2}{\pi^2} \left(\frac{a}{l_0}\right)^3
\left(\frac{1}{\hbar\omega_g}\right)^3
\sin^2{\left(\frac{ql_0}{2 \epsilon_F^g} \right)}.
\end{equation}
Here, $ \epsilon = E/(\hbar \omega_g) $ and 
$ \epsilon_{_F}^g = E_{F}^g/(\hbar\omega_g)$.

The fluctuation in the DOS for electrically modulated graphene \cite{khan} 
is proportional to $\cos^2\left(ql_0\epsilon_F^g - \pi/4\right) $.
The appearance of the square of the sine term in Eq. (\ref{dm}) instead of 
square of cosine is the reason of definite phase differences in
the fluctuation of all the thermodynamic quantities.

Using the two Eqs. (\ref{du}) and (\ref{dm}) separately, 
we get an approximate analytical expression of the free energy and 
it's fluctuation.
Our aim is to study the magnetic 
modulation effect in compare to the electrical modulation in graphene \cite{khan}. 
The change in the free energy due to the magnetic modulation can 
be expressed as

\begin{eqnarray}
\frac{F_m}{A} & = & - (k_{_B} T)\hbar\omega_g \int_0^\infty D_m(\epsilon)
\ln\left[ 1 + \exp\left(\frac{\mu - \hbar\omega_g\epsilon}{k_{_B} T }\right)\right]
d\epsilon \nonumber \\
& = & - \Omega_m \frac{1}{\pi l_0^2}
4(\epsilon_{_F}^g)^4(k_{_B}T)\hbar\omega_g
\sin^2\left(ql_0 \epsilon_{_F}^g -\frac{\pi}{4}\right)
\nonumber\\ 
& \times & \int_0^\infty \cos\left(\pi \epsilon^2\right)
\ln\left[1+\exp\left(\frac{\mu-\hbar\omega_g\epsilon}{k_{_B} T}\right)\right]
d\epsilon. \nonumber
\end{eqnarray}
Under the assumption of very low temperature the above integration results to
\begin{eqnarray}
\frac{F_m}{A}  & = -& \frac{\Omega_m}{\pi l_0^2}\frac{(\hbar\omega_g\epsilon_F^g)^4}{\pi} 
\Big\{\sin\left(\pi (\epsilon_{_F}^g)^2\right)
\nonumber\\ 
& - & \left(1-\frac{T/T_g^{dHvA}}{\sinh(T/T_g^{dHvA})}\right)
\frac{1}{2}\frac{\cos\left(\pi (\epsilon_{_F}^g)^2\right)}
{\left(\pi (\epsilon_{_F}^g)^2\right)}
\Big\} \nonumber\\ 
& \times & \sin^2\left(ql_0 \epsilon_{_F}^g - 
\frac{\pi}{4}\right),\label{free}
\end{eqnarray}
where $T_g^{dHvA} = (\hbar\omega_g)/(2 \pi^2 k_{_B} \epsilon_{_F}^g)$ 
is the critical
temperature for the dHvA-type oscillations in graphene.
The ratio of amplitude of the free energy 
fluctuations of the magnetically and electrically modulated graphene is 
\begin{eqnarray}
\frac{\lambda_m^g}{\lambda_e^g} \approx
\Big(\frac{V_m^g}{V_e^g}\Big)^2.
\end{eqnarray}
Here, $\lambda_m^g $ and $\lambda_e^g $ are the amplitudes of free energy 
fluctuation for magnetically and electrically modulated graphene, respectively.
The expression of $\lambda_e^g $ is taken from the Ref. \cite{pat}.
But in the case of conventional 2DEG this ratio has been calculated 
in Ref. \cite{stewart} and it is given by
\begin{equation}
\frac{\gamma_m^{2d}}{\gamma_e^{2d}} = \frac{1}{2\pi^2} 
\left(\frac{E_F^{2d}}{\epsilon_a}\right) \left(\frac{V_m^{2d}}{V_e^{2d}} \right)^2,  
\end{equation}
where $\epsilon_a = \hbar^2/(m^*a^2)$, $ \gamma_m^{2d} $ and $ \gamma_e^{2d} $ are amplitudes
of the free energy fluctuation for magnetic and electric modulation, respectively.
The ratio of the amplitudes of different modulation cases in graphene behave differently 
from that of the 2DEG system.

Now we compare our result with the conventional 2DEG modulated magnetically. 
In this two case the ratio of amplitudes of the free energy fluctuations 
is given by
\begin{eqnarray}
\frac{\gamma_m^{2d}}{\lambda_m^{g}} & = & \left(\frac{V_m^{2d}}{V_m^g}\right)^2
\frac{\pi}{(\beta E_F^g)^2} \left(\frac{a}{l_0}\right)^2 
\sqrt{\frac{ E_F^{2d}}{2 \hbar\omega_0}}
\left(\frac{\omega_g}{\omega_0}\right)^3,
\end{eqnarray}
where $E_F^g$ is the Fermi energy of graphene whereas $E_F^{2d}$ is for 
conventional 2DEG. Using $B_0 = 0.1$ T, $T=2$ K, $ a=382$ nm, 
$\beta E_F^g=521$, 
$\omega_g/\omega_0 = 35 $, we get
\begin{equation}
\frac{\gamma_m^{2d}}{\lambda_m^{g}}\sim10^2\left(\frac{V_m^{2d}}{V_m^g}\right)^2.
\end{equation}
This equation leads to $\lambda_m^{g} \sim 4.7 \lambda^{2d}$ when $ B_{1}=0.02$ T. 
In graphene, the amplitude of free energy fluctuation in the Weiss-type oscillation 
is higher and in the same phase in compare to the conventional 2DEG.

\section{Electric and Magnetic Modulations}
In this section, we study how the thermodynamic properties 
discussed in the previous sections are changed in presence of 
an additional electric modulation with the same period.
We consider two different cases: when both the modulations are 
in-phase and that are $\pi/2 $ out-of-phase.
The fluctuation in thermodynamic quantities like chemical potential 
and the Helmholtz free energy are calculated numerically. 
Other thermodynamic quantities can easily 
be obtained by taking a suitable numerical derivative of the Helmholtz 
free energy fluctuation.  

{\bf In-phase modulations:}
We consider a weak electric modulation described by the 
periodic potential $ U(x) = V_e^g \cos(qx) $, which is
in-phase with the magnetic modulation 
$ {\bf B}_1(x) = B_1 \cos(qx) \hat z$.   
When both the modulations are in the same phase, 
the total energy correction for graphene in a weak magnetic field 
can be written as 
\begin{eqnarray}
\Delta E^g = \sqrt{f_m^2+f_e^2}\sin(2\sqrt{nu} - 
\frac{\pi}{4}+\delta_{i}^{g}) \cos(qx_0),
\end{eqnarray}
where $f_m = (2V_m^g/\sqrt{\pi}) (n/u^3)^{1/4}
\sin{(\sqrt{u/4n})}$,
$f_e = (V_e^g/\sqrt{\pi}) (1/nu)^{1/4} \cos(\sqrt{u/4n})$ and 
$\delta_{i}^g = \tan^{-1}(f_m/f_e)$.
The flat-band condition at the Fermi energy gives the positions 
of the minima in the free energy fluctuation as 
$ B_j = 2p_{_F} B_a/(j+\frac{1}{4}-\frac{\delta_{i}^g}{\pi})$.
Here, $j$ is an integer, $p_{_F} =ak_{_F} $ is a dimensionless momentum
and $B_a = \hbar/(ea^2)$ is the characteristic
magnetic field. In this case, $\delta_{i}^{g} = \pi/4$ and
then $B_j = 2p_{_F}B_a/j$.

Similarly for conventional 2DEG, the total energy correction 
in the low magnetic field can be written as 
\begin{equation}
\Delta E^{2d} \simeq \sqrt{w_m^2 + w_e^2} 
\sin(2\sqrt{nu}-\frac{\pi}{4} + \delta_{i}^{2d})
\cos(qx_0),
\end{equation}
where $w_m \simeq V_m^{2d}\sqrt{n/(\pi u \sqrt{nu}}) $,
$w_e \simeq V_e^{2d}/(\sqrt{\pi \sqrt{nu}}) $ and 
$\delta_{i}^{2d}=\tan^{-1}(w_m/w_e)$.
The flat-band condition at the Fermi energy is now 
$ B_j = 2p_{_F} B_a/(j+ 1/4 - \tan^{-1}\{V_m^{2d} p_{_F}/( 2\pi V_e^{2d})\})$.

In figures 8 and 9, we plot the chemical potential and free energy
fluctuations in presence of both electric and magnetic modulations, respectively.
The upper panel shows the thermodynamic fluctuations of graphene (solid) and
2DEG (dashed) when the modulations are in-phase and the lower panel shows
the thermodynamic fluctuations of graphene (solid) and
2DEG (dashed) when the modulations are out-of-phase. 

\begin{figure}[ht]
\begin{center}\leavevmode
\includegraphics[width=98mm]{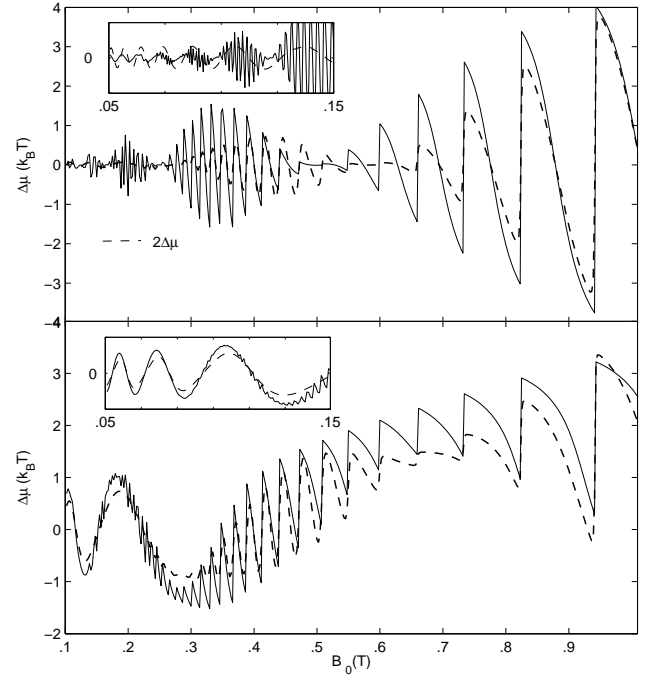}
\caption{Plots of the fluctuation in chemical potential vs magnetic field.} 
\label{Fig7}
\end{center}
\end{figure}

{\bf Out-of-phase modulation:}
We consider the same electric modulation $ U(x) = V_e \cos (qx) $
and assume magnetic modulation is given by 
$ {\bf B}_1(x) =  B_1 \sin (qx) \hat z $ so that the two modulations are
$\pi/2$ out-of-phase. To first-order in 
$V_e $ and $ B_1$, the total energy correction for graphene
can be written as
\begin{eqnarray}
\Delta E^g &=& f_m\sqrt{1+\{(\frac{f_e}{f_m})^2-1\}\cos^2(2\sqrt{nu}-1)}\nonumber\\
&\times& \sin(qx_0 + \delta_{o}^g),
\end{eqnarray}
where 
$ \tan (\delta_{o}^g) = (f_m/f_e) \tan(2\sqrt{nu} - \pi/4) $.   
Similarly, the total energy correction for 2DEG is written as
\begin{eqnarray}
\Delta E^{2d} &=& w_m\sqrt{1+\{(\frac{w_e}{w_m})^2-1\}
\cos^2(2\sqrt{nu}-1)}\nonumber\\
& \times & \sin(qx_0 + \delta_{o}^{2d}),
\end{eqnarray}
where $ \tan (\delta_{o}^{2d}) = (w_m /w_e) 
\tan (2 \sqrt{nu} - \pi/4)$.
In the case of graphene, though the band width is almost constant at the Fermi energy, 
but the magnetic field dependent phase factor $\delta^g_{o}$ plays an 
important role in the fluctuation of 
the thermodynamic quantities. This phase factor produces the Weiss-type
oscillations in the thermodynamic properties at low magnetic field. 
In conventional 2DEG, the band width oscillates with the magnetic
field when $ w_e \neq w_m $ in our case but the Weiss-type oscillation 
is due to the magnetic field dependent phase factor 
$\delta_{o}^{2d}$ in the total energy correction.
When $ w_e = w_m $, the band width becomes non-oscillatory, but the 
Weiss-type oscillation still exist due to the magnetic field dependent phase
factor. 
It is interesting to contrast our result with the results of 
the Weiss oscillations
in the conductivity in presence of both the modulations \cite{fm,ijmp}.
The Weiss oscillation in conductivity is suppressed when the modulations are
out-of-phase. 
On the other hand, the Weiss-type oscillations in the thermodynamic 
properties enhanced when the modulations are out-of-phase. 
In out-of-phase case, the Weiss-type oscillation 
in the thermodynamic quantities is due to the magnetic field
dependent phase factor in the energy correction.
\\
\begin{figure}[ht]
\begin{center}\leavevmode
\includegraphics[width=98mm]{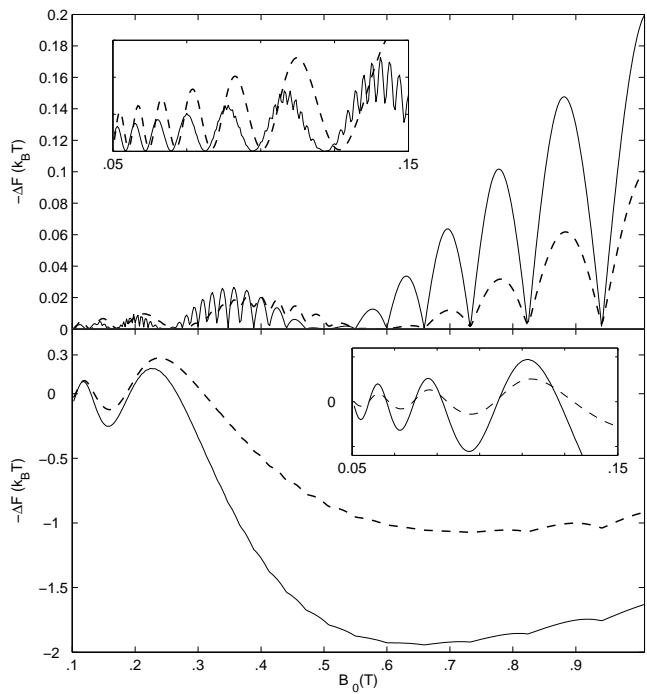}
\caption{Plots of the fluctuation in the Helmholtz free energy vs magnetic field.} 
\label{Fig8}
\end{center}
\end{figure}
\\
\section{Summary}
We have studied the effect of magnetic modulation on thermodynamic 
properties of a graphene sheet. The results of magnetically modulated 
graphene are compared with electrically modulated graphene and magnetically 
modulated conventional 2DEG. It is observed that in the case of 
magnetically modulated graphene, a definite phase difference is 
appeared in the Weiss-type oscillation in compare to the electrically 
modulated graphene for all thermodynamic quantities. 
But when we compare our results with magnetically modulated conventional 2DEG,
the amplitude of the fluctuations is found to be higher in graphene than 2DEG,
but the phases remain same.
We calculate the DOS and the Helmholtz free energy analytically.
and explain the origin of the this phase difference in the 
Weiss-type oscillation.
The enhancement of the fluctuation in magnetically modulated graphene in compare
to the 2DEG and the definite phase difference between magnetically and electrically
modulated graphene are explained by using the asymptotic results of the DOS in general
and the Helmholtz free energy in particular.

We have also studied the thermodynamic properties like 
chemical potential and the Helmholtz free energy of graphene and conventional
2DEG in presence of both magnetic and electric modulations with the same period.
The combined effect of both modulation does not modify the nature of
the Weiss-type oscillation when they are in the same phase 
except a finite phase shift in the fluctuation. 
The effect of the out-of-phase modulations on thermodynamic fluctuations is
different than that of in-phase modulation. We found large amplitude
Weiss-type oscillation at very low magnetic field even when
bandwidth becomes non-oscillatory. 
For conventional 2DEG, effect of the out-of-phase modulation
remains same as graphene though the bandwidth of 2DEG is oscillatory. 
This high amplitude Weiss-type
oscillation is due to the magnetic field dependent phase factor which
plays an important role here, unlike the Weiss oscillation in 
electrical transport properties where oscillation washed out 
for the same amplitude of energy correction due to two modulations.


\begin{acknowledgements}
This work is financially supported by C.S.I.R., Govt. of India under the 
grant CSIR-JRF-09/092(0687) 2009/EMR-1 F-O746.
\end{acknowledgements}

\begin{appendix}

\section{}
Here we calculate the asymptotic expression of the DOS of 
graphene in presence of a modulated magnetic field:   
$ {\bf B} = [B_{0} +  B_{1} \cos{(qx)}] \hat z$.
The total energy upto the first-order in $V_m^g$ is
$ E_{n,k_y}^{g,m} = \hbar \omega_g \sqrt{2n}+ G_n\cos{(qx_0)} $,
where 
$ G_n = V_m^g \sqrt{n/u} e^{-u/2} [L_{n-1}(u)-L_n(u)] $
and $u=q^2l_0^2/2$.
Using $ e^{-u/2} L_n(u) \simeq (\pi\sqrt{nu})^{-1/2}\cos[2\sqrt{nu}-
\pi/4]$ for higher values of $n$,
we get the asymptotic expression of $G_n$ as
\begin{eqnarray}
G_n & = & \frac{2V_m^g}{\sqrt{\pi}} \left(\frac{n}{u^3}\right)^{1/4}
\sin{\left(\sqrt{\frac{u}{4n}}\right)}\sin\Big\{2\sqrt{nu}-
\frac{\pi}{4}\Big\} \nonumber \\ 
& = & \frac{2V_m^g}{\sqrt{\pi}}
\left[\left(\frac{2E}{ ql_0\hbar\omega_g}\right)^{1/2}
\frac{1}{ql_0}\right] \nonumber \\ 
& \times & \sin{\left(\frac{ql_0\hbar\omega_g}{2E}\right)}
\sin\Big\{ql_0\frac{E}{\hbar\omega_g} - \frac{\pi}{4}\Big\}.
\end{eqnarray}
Now we use impurity broadened Landau levels in limiting case. 
The self-energy can be written as\cite{eco}
\begin{equation}
\Sigma^-(E) = \Gamma_0^2 \sum_n \int_0^a 
\frac{dx_0}{a}\frac{1}{E-E_{n,k_y}^{g,m} - \Sigma^-(E)},
\end{equation}
where $\Gamma_0$ is the broadening of the Landau levels 
due to impurities. By determining the imaginary part of the self-energy 
we can get the DOS through
\begin{equation}
D(E) = \Im \left[\frac{\Sigma^-(E)}{\pi^2 l_0^2 \Gamma_0^2}\right] 
\label{density}.
\end{equation}
By using residue theorem we get
\begin{eqnarray}
\Sigma^-(E) & = & 2\pi\Gamma_0^2\int_0^a \frac{dx_0}{a}
\frac{E-\Sigma^-(E)-G_n\cos(qx_0)}{(\sqrt{2}\hbar\omega_g)^2} \nonumber \\
& \times & \cot\left[\frac{\pi}{(\sqrt{2}\hbar\omega_g)^2}\{E-\Sigma^-(E)-G_n\cos{(qx_0)}\}^2
\right] \nonumber \\ 
& \simeq & \frac{\pi\Gamma_0^2E}{(\hbar\omega_g)^2}\int_0^a\frac{dx_0}{a}
\times \nonumber \\ 
& \times & \cot\left[\frac{\pi E}{(\sqrt{2}\hbar\omega_g)^2}\{E-2\Sigma^-(E)-2G_n\cos{(qx_0)}\}
\right].\nonumber
\end{eqnarray}
Separating $\Sigma^-(E)$ into real and imaginary parts as
\begin{eqnarray}
\Sigma^-(E) & = & \Delta(E) + i\frac{\Gamma(E)}{2} \nonumber \\ 
& = & \frac{\pi\Gamma_0^2 E}{(\hbar\omega_g)^2}
\int_0^a \frac{dx_0}{a} \Big[\frac{\sin{u}+i\sinh{v}}{\cosh{v}-\cos{u}}\Big] 
\label{sin},
\end{eqnarray}
where 
\begin{equation}
u = \frac{\pi E}{(\hbar\omega_g)^2}[E-2\Delta(E)-2G_n\cos{(qx_0)}], \nonumber\\
\end{equation}
and
$ v = \frac{\pi E}{(\hbar\omega_g)^2}\Gamma(E)$.
In the limit of large collisional broadening, 
$\pi \Gamma \gg \hbar \omega_g$, after expanding the hyperbolic
term with respect to the small quantity
exp$(-v)$ up to first-order one obtains
\begin{eqnarray}
\frac{\Gamma(E)}{2} & = & \frac{\pi\Gamma_0^2E}{(\hbar\omega_g)^2}
\Big[1+2\exp\left\{-\frac{\pi E\Gamma}{(\hbar\omega_g)^2}\right\} \nonumber \\
& \times & \int_0^a\frac{dx_0}{a}\cos \Big\{\frac{\pi E}{(\hbar\omega_g)^2}
(E-2G_n\cos(qx_0)) \Big\}\Big] \label{img} \nonumber\\.
\end{eqnarray}
After first iteration, we have
$ \Gamma(E)/2 = \pi\Gamma_0^2 E/(\hbar\omega_g)^2$,
and then putting it back into Eq. (\ref{img}) and using $\epsilon=E/(\hbar\omega_g)$, we get
\begin{eqnarray}
\frac{\Gamma(\epsilon)}{2} & \backsimeq & \frac{\pi\Gamma_0^2\epsilon}{(\hbar\omega_g)}
\Big[1+2\exp\left\{-2\Big(\frac{\pi \epsilon\Gamma_0}{(\hbar\omega_g)}\Big)^2\right\}
\nonumber\\&\times&\cos\left(\pi \epsilon^2\right)
\nonumber\\ & \times &
\Big\{1-\Omega_m (\epsilon\hbar\omega_g)^3\sin^2\left(ql_0\epsilon
-\frac{\pi}{4}\right)\Big\}\Big], \nonumber\\ \label{sigma}
\end{eqnarray}
where 
\begin{equation}
\Omega_m = \frac{(V_m^g)^2}{\pi^2} \left(\frac{a}{l_0}\right)^3
\left(\frac{1}{\hbar\omega_g}\right)^5
\sin^2{\left(\frac{ql_0}{2\epsilon}\right)}.
\end{equation}
Using Eqs. (\ref{density}) and (\ref{sigma}), we get the 
density of states $D(\epsilon)$ as given in Eqs. (\ref{du}) and (\ref{dm}).

\end{appendix}

\end{document}